\documentclass[aps,pra,reprint,amsmath,amssymb,superscriptaddress,longbibliography]{revtex4-2}
\usepackage[american]{babel}
\addto\captionsamerican{}
\usepackage{amsthm}
\usepackage{graphicx,bm}
\usepackage[colorlinks=true,linkcolor=blue,citecolor=blue,urlcolor=blue]{hyperref}
\hypersetup{pdftitle={Quantifying extractable imaginarity with qubit triangles},pdfsubject={Imaginarity extraction under real operations},pdfauthor={Meng-Li Guo, Chen-Xin Xu, Wen Xu, and Shao-Ming Fei}}
\theoremstyle{definition}
\newtheorem{theorem}{Theorem}
\newtheorem{lemma}{Lemma}
\newtheorem{proposition}{Proposition}
\newtheorem{corollary}{Corollary}
\newcommand{\Tr}{\operatorname{Tr}}
\newcommand{\supp}{\operatorname{supp}}
\newcommand{\rank}{\operatorname{rank}}
\newcommand{\conv}{\operatorname{conv}}
\newcommand{\ii}{\mathrm{i}}
\newcommand{\Dtri}{\mathcal D_{\triangle}}
\newcommand{\gtri}{g_{\triangle}}
\newcommand{\RR}{\mathbb R}

\newcommand{\id}{\mathbb I}
\newcommand{\ket}[1]{\lvert #1\rangle}
\newcommand{\bra}[1]{\langle #1\rvert}
\newcommand{\norm}[1]{\lVert #1\rVert}

\begin{document}
\title{Quantifying extractable imaginarity with qubit triangles}
\author{Meng-Li Guo}
\email{guomengli2020@163.com}
\affiliation{School of Science, East China University of Technology, Nanchang 330006, China}
\author{Chen-Xin Xu}
\email{chenxin200401@163.com}
\affiliation{School of Science, East China University of Technology, Nanchang 330006, China}
\author{Wen Xu}
\affiliation{School of Science, East China University of Technology, Nanchang 330006, China}
\author{Shao-Ming Fei}
\email{feishm@cnu.edu.cn}
\affiliation{School of Mathematical Sciences, Capital Normal University, Beijing 100048, China}
\begin{abstract}
How much imaginarity does a state hold, and how much can real operations extract? We connect these questions through a triangle assigned to each qubit. Its base $R$ is the robustness of imaginarity, and its slope $\kappa$ the largest robustness attainable by real filtering. In a source-comparison task, a receiver decides whether two draws selected a state or its conjugate; the optimal mean success probability is $2/3+R^2\kappa/3$. The normalized area $R^2\kappa$ defines a convex measure that cannot increase on average under real instruments and vanishes exactly on real states. In any finite dimension, we maximize the mean area of the extracted qubits, the minimal strongly monotone extension of the qubit measure. Two eight-dimensional states agree in robustness, conditional robustness, and optimal quadratic yield, yet their areas differ by $3/1000$, excluding a deterministic real conversion. A real projection attains the optimal quadratic yield, the Bures quantum $\chi^2$ divergence between the input and its real part, giving bounds $J^2/R\le\Dtri\le\kappa J$. The projection preserves the midpoint quantum Fisher information of a conjugate mixture when the label is retained. Exact areas follow for commuting real and imaginary parts and for pure states mixed with white noise.
\end{abstract}
\maketitle

\section{Introduction}\label{sec:intro}
Diagonal and circular polarizations have the same populations in the horizontal and vertical basis. A real rotation takes diagonal polarization to a basis state, whereas circular polarization retains an imaginary off-diagonal entry under any such rotation. A quarter-wave plate supplies the phase shift that allows a linear polarizer to distinguish the two circular polarizations~\cite{WuPRA2021}. This illustrates how the available controls determine which part of coherence is associated with imaginarity~\cite{GuoCoherence2026}. A related question is whether an entire experiment admits a real description. Restricting pure-state decompositions to real vectors can increase the entanglement of formation of a pair of rebits~\cite{Caves2001}. Enlarging the real Hilbert space, in turn, permits simulations of complex systems~\cite{McKague2009}. When a network has independent sources, the simulation must also respect their independence. In this setting real and complex theories can give different predictions~\cite{Renou2021}, a distinction explored in optical and superconducting experiments~\cite{LiReal2022,ChenReal2022}.

The resource theory of imaginarity poses the question within ordinary complex quantum mechanics~\cite{Hickey2018,Chitambar2019}. A reference basis is fixed by the laboratory controls. Free states have real density matrices in this basis, and free operations admit real Kraus representations. Any individual state is real in its own eigenbasis, but implementing that change of basis can itself require control of a relative phase. For several preparations a different question arises when the same basis must serve for all of them. Bargmann invariants can reveal imaginarity in collections of states~\cite{LiTan2025}. For qubits, Li, Wagner, and Zhang characterize when a single basis change makes every state in a collection real~\cite{Li2026}. For a single preparation in a fixed basis, restricting the laboratory controls gives operational content to imaginarity in state conversion and discrimination~\cite{WuPRL2021,WuPRA2021} and in distributed protocols~\cite{Wu2024}.

Different measures of imaginarity answer different questions about a preparation. Robustness asks how much mixing with another state is needed to reach the set of real states. Geometric imaginarity quantifies proximity to that set, and both measures admit explicit expressions~\cite{WuPRA2021}. Other quantifiers use entropy or construct mixed-state measures from their values on pure-state ensembles~\cite{Xue2021,Xu2024,DuBai2025,Guan2026}. State-conversion results specify which targets are reachable under real operations and when a probabilistic transformation can succeed~\cite{WuPRA2021,Kondra2023,Guo2025}. In an estimation experiment the relevant comparison is between nearby outputs of a parameterized preparation or channel. Quantum Fisher information expresses their local distinguishability in terms of a metric on states~\cite{Petz1996,Braunstein1994}. For the channel encoding studied by Lu et al.~\cite{Lu2026}, even qubit probes of equal robustness can have different sensitivities.

One way to test what imaginarity contributes to a preparation is to compare two source choices. Suppose a source emits a known state or its complex conjugate with prescribed probabilities. Given two independent preparations, can a receiver tell whether the same choice was made twice? For two known states and a fixed prior, minimum-error comparison reduces to optimal discrimination on each system~\cite{Hayashi2018Comparison}. A sufficiently strong prior can already favor one answer, even before either system is measured. At fixed robustness of imaginarity, over what range of biases does a measurement improve the guess, and how large is its average gain?

We show that the largest robustness attainable by real filtering gives the precise bias at which a measurement ceases to improve source comparison. Averaging over a uniformly distributed bias identifies the gain with one third of the normalized qubit triangle area. For a higher-dimensional input, an instrument can leave a qubit and a classical record on each run. The largest average area of these outputs gives the pointwise smallest strongly monotone extension of the qubit measure in the framework of Gour and Tomamichel~\cite{Gour2020}. We bound this area by solving the corresponding squared-robustness problem with an explicit real projection. Two eight-dimensional preparations agree in their robustness, conditional robustness, and optimal quadratic yield, yet the area excludes a deterministic real conversion in one direction.

\section{Triangular imaginarity}\label{sec:geometry}
\subsection{Real operations and conditional extraction}
Fix an orthonormal basis $\{\ket j\}_{j=1}^d$ for transposition and complex conjugation. A density operator then has the form
\begin{equation*}
 \rho=A+\ii B,\quad A=A^T\ge0,\quad B=-B^T,\quad \Tr A=1.
\end{equation*}
Here $A$ and $B$ are real, and $B=0$ precisely for the free states. We describe a real instrument by real Kraus operators $K_\mu$ with $\sum_\mu K_\mu^TK_\mu=\id$. Outcome $\mu$ occurs with probability $p_\mu$ and leaves the conditional state $\rho_\mu$, where
\begin{equation*}
 p_\mu=\Tr(K_\mu\rho K_\mu^T),\qquad
 \rho_\mu=K_\mu\rho K_\mu^T/p_\mu.
\end{equation*}
We sum only over outcomes of nonzero probability. When the outcome is recorded, strong monotonicity bounds the average resource of the conditional states. Convexity accounts for partial or complete loss of the record. These requirements also underlie measures of entanglement~\cite{Vidal2000} and coherence~\cite{Baumgratz2014,Streltsov2017}.

Robustness quantifies the amount of mixing required to make a state free. Originally developed for entanglement~\cite{VidalTarrach1999,Steiner2003} and also studied for coherence~\cite{Napoli2016}, robustness measures admit convex formulations and operational interpretations in discrimination tasks~\cite{Regula2018,TakagiRegula2019}. For imaginarity, the robustness reduces to~\cite{WuPRA2021}
\begin{equation}
 R(\rho)=\tfrac12\norm{\rho-\rho^T}_1=\norm B_1.
 \label{eq:R}
\end{equation}
To determine how much robustness can be obtained in a single successful filtering outcome, consider the quantity
\begin{equation}
 \kappa(\rho)=\norm{A^{-\tfrac12}(\ii B)A^{-\tfrac12}}_\infty.
 \label{eq:kappa}
\end{equation}
Inverses throughout are taken on $\supp A$ and set to zero on its orthogonal complement. Positivity gives $\supp B\subseteq\supp A$, and Eq.~\eqref{eq:kappa} yields
\begin{equation}
 -\kappa A\le\ii B\le\kappa A,\qquad 0\le R\le\kappa\le1.
 \label{eq:order}
\end{equation}
For a real state, $\kappa=0$. A real completely positive map preserves the order in Eq.~\eqref{eq:order}. Applying the map for outcome $\mu$ therefore gives
$\kappa(\rho_\mu)\le\kappa(\rho)$. This bound can be attained by a qubit output. A real filter first whitens $A$ on its support and then selects a two-dimensional canonical block on which $A^{-\tfrac12}(\ii B)A^{-\tfrac12}$ attains its norm. For a nonreal input, the selected block occurs with nonzero probability and has robustness $\kappa(\rho)$ (Appendix~\ref{app:support}). Thus $\kappa$ gives the largest conditional qubit robustness when the success probability is unrestricted.

For a qubit $\tau=\tfrac12(\id+\bm r\cdot\bm\sigma)$, let
\begin{equation*}
 u=\sqrt{r_x^2+r_z^2},\qquad a=\sqrt{1-u^2}.
\end{equation*}
Equations~\eqref{eq:R} and~\eqref{eq:kappa} give
\begin{equation*}
 R=|r_y|,\qquad \kappa=R/a.
\end{equation*}
At $a=0$, positivity forces $R=0$, and we set $\kappa=0$. The criterion of Ref.~\cite{WuPRA2021} for deterministic qubit conversion under real operations now reads
\begin{equation}
 \tau\longrightarrow\omega
 \quad\Longleftrightarrow\quad
 \begin{cases}
 R(\tau)\ge R(\omega),\\
 \kappa(\tau)\ge\kappa(\omega).
 \end{cases}
 \label{eq:conversion}
\end{equation}

\subsection{The qubit triangle}
We use the two parameters in Eq.~\eqref{eq:conversion} to associate a right triangle with each qubit state,
\begin{equation}
 T(\tau)=\conv\{(0,0),(R,0),(R,R\kappa)\}.
 \label{eq:triangle}
\end{equation}
\begin{figure}[t]
 \centering
 \includegraphics[width=\columnwidth]{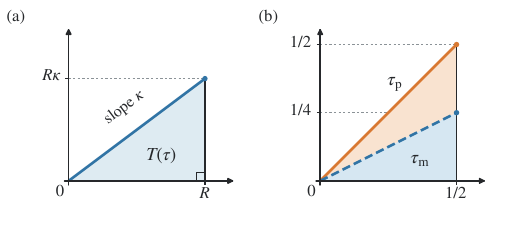}
 \caption{Qubit triangles. (a) The base $R$ and height $R\kappa$ give a normalized area $2\operatorname{Area}T(\tau)=R^2\kappa$. (b) The states in Eq.~\eqref{eq:qubitexample} have the same base $R=1/2$. The mixed state (blue dashed line) and pure state (orange solid line) have normalized areas $1/8$ and $1/4$, respectively. Under the same source-comparison test the optimal mean success probabilities are $17/24$ and $3/4$, respectively [Eq.~\eqref{eq:comparison}].}
 \label{fig:one}
\end{figure}

In Fig.~\ref{fig:one}(a), the base length gives the input robustness $R$, while the slope of the hypotenuse gives its maximum conditional value $\kappa$. One triangle contains another precisely when both its base and its slope are at least as large. Thus $T(\omega)\subseteq T(\tau)$ is the geometric form of Eq.~\eqref{eq:conversion}. We define the triangular imaginarity of a qubit by its normalized area,
\begin{equation}
 \gtri(\tau)=2\operatorname{Area}T(\tau)
 =R^2\kappa=\frac{|r_y|^3}{\sqrt{1-r_x^2-r_z^2}}.
 \label{eq:g}
\end{equation}
On real states we set $\gtri=0$. Since $0\le\gtri\le R^2$, the area extends continuously to the entire Bloch ball. For nonreal pure states $\kappa=1$, giving $\gtri=R^2$. Along the $y$ axis ($r_x=r_z=0$), the same formula gives $\gtri=R^3$.

To relate the triangle to a measurement, consider the following source comparison. Two sources emit the known qubit states $\tau$ and $\tau^T$ in the fixed reference basis. In each round a bias $t$ is drawn uniformly from $[0,1]$ and announced. Conditional on this bias, two independent draws select the first source with probability $p=(1+t)/2$ and the second with probability $1-p$. A receiver measures the two qubits and guesses whether the same source was chosen twice. The measurement can depend on $t$ and may be any quantum measurement.

\begin{corollary}\label{cor:comparison}
For every qubit state $\tau$, the optimal success probability averaged over the announced bias is
\begin{equation}
 \overline P_{\rm cmp}^{\,\rm opt}(\tau)=\frac23+\frac{\gtri(\tau)}3.
 \label{eq:comparison}
\end{equation}
At a fixed bias, measuring the qubits improves on guessing from the prior alone precisely when $t<\kappa(\tau)$.
\end{corollary}
\begin{proof}
Put $D_t=p\tau-(1-p)\tau^T$. For two independent draws, the prior-weighted same-source and different-source states differ by $D_t\otimes D_t$, as in the general comparison result of Hayashi \textit{et al.}~\cite{Hayashi2018Comparison}. The Helstrom formula~\cite{Helstrom1976} and multiplicativity of the trace norm give $P_{\rm cmp}^{\rm opt}(t)=(1+\norm{D_t}_1^2)/2$. Measuring the sign of $D_t$ on each qubit and comparing the two outcomes attains this value, with either sign assigned on the kernel. Guessing that the sources agree gives $p^2+(1-p)^2=(1+t^2)/2$ from the prior alone.

The eigenvalues of $D_t$ are $(t\pm\sqrt{R^2+u^2t^2})/2$. Thus $\norm{D_t}_1^2-t^2=\max\{0,R^2-a^2t^2\}$. This excess vanishes for a real state. For a nonreal state it is positive for $t<\kappa=R/a$ and zero for $t\ge\kappa$. Averaging the prior success probability gives $2/3$, and the mean gain is
\begin{equation*}
 \frac12\int_0^\kappa (R^2-a^2t^2)\,dt
 =\frac{R^2\kappa}{3}=\frac{\gtri(\tau)}{3}.\qedhere
\end{equation*}
\end{proof}

Triangle inclusion implies that deterministic conversion cannot increase the area. Selective operations require a separate argument. For example, the concurrence fill of a tripartite entanglement triangle~\cite{XieEberly2021} can increase on average under local operations and classical communication~\cite{Ge2023}. For the triangle in Eq.~\eqref{eq:triangle}, we obtain the following bound.

\begin{lemma}\label{lem:qubit}
The function $\gtri$ is faithful and convex. For every real qubit instrument,
\begin{equation}
 \sum_\mu p_\mu\gtri(\tau_\mu)\le\gtri(\tau).
 \label{eq:gstrong}
\end{equation}
\end{lemma}
For qubits, invertible real filters preserve $\kappa$ and rank-one filters produce real outputs. Appendix~\ref{app:qubit} uses this distinction and a determinant inequality to bound the average area.

To see what the slope contributes at fixed robustness, consider the mixed state $\tau_{\rm m}$ and the pure state $\tau_{\rm p}$ given by
\begin{align}
 \tau_{\rm m}&=\tfrac12(\id+\tfrac12\sigma_y),\nonumber\\
 \tau_{\rm p}&=\tfrac12(\id+\tfrac12\sigma_y+\tfrac{\sqrt3}{2}\sigma_z).
 \label{eq:qubitexample}
\end{align}
Both states have $R=1/2$ and hence the same geometric imaginarity $I_G=(1-\sqrt{1-R^2})/2$~\cite{WuPRA2021}. The largest robustness attainable by filtering is $1/2$ for $\tau_{\rm m}$ and $1$ for $\tau_{\rm p}$. At zero bias both preparations give the same comparison success probability. With increasing bias, the mixed state ceases to improve on the prior at $t=1/2$, whereas the pure state retains an advantage for every $t<1$. Their normalized areas, $1/8$ and $1/4$ [Fig.~\ref{fig:one}(b)], give a difference of $1/24$ in mean success probability by Eq.~\eqref{eq:comparison}. Up to a real orthogonal rotation, Lu et al.~\cite{Lu2026} studied the same pair in a channel-estimation model.

\subsection{Extension by qubit extraction}
For a $d$-dimensional input, consider a real instrument with qubit outputs and a classical record of the outcome. Its finitely many Kraus operators $K_\mu\in\RR^{2\times d}$ satisfy $\sum_\mu K_\mu^TK_\mu=\id_d$. We define the extractable triangular imaginarity as the supremum of the average qubit area over all such instruments,
\begin{equation}
 \Dtri(\rho)=\sup_{\{K_\mu\}}\sum_\mu p_\mu\gtri(\rho_\mu).
 \label{eq:Ddef}
\end{equation}
Proposition~\ref{prop:finite} shows that this supremum is attained. Equation~\eqref{eq:Ddef} assigns the triangle to each extracted qubit, so the definition applies even when two parameters do not describe the conversion order of the input state.

\begin{theorem}\label{thm:D}
In every finite dimension, $\Dtri$ is faithful, convex, and strongly monotone under real instruments, with $0\le\Dtri\le1$. It agrees with $\gtri$ on qubits and is invariant under adjoining real ancillary states. For a direct sum on real orthogonal subspaces, $\Dtri(\bigoplus_jq_j\rho_j)=\sum_jq_j\Dtri(\rho_j)$ whenever $q_j\ge0$ and $\sum_jq_j=1$. Every strong monotone $M$ defined across finite dimensions that agrees with $\gtri$ on qubits satisfies $M(\rho)\ge\Dtri(\rho)$ for every state $\rho$.
\end{theorem}
\begin{proof}
For a fixed instrument, $p_\mu\gtri(\rho_\mu)$ is the perspective of the convex function $\gtri$~\cite{Boyd2004} evaluated on the unnormalized output. The sum over outcomes is therefore convex in the input, as is its supremum over instruments. To prove strong monotonicity, follow each outcome of a real instrument by a qubit extraction arbitrarily close to optimal for that outcome. The composition is admissible in Eq.~\eqref{eq:Ddef}, so its average area is bounded by $\Dtri(\rho)$. For a qubit input, Lemma~\ref{lem:qubit} bounds this average by $\gtri$, and the identity instrument attains the bound.

Strong monotonicity of $R$ and the inequality $\gtri\le R$ give $\Dtri(\rho)\le R(\rho)\le1$. A real input produces only real outputs. If $B\ne0$, choose $j,k$ with $B_{jk}\ne0$ and project onto the corresponding coordinate plane. This outcome has positive probability and positive area, proving faithfulness. A real ancilla can be added and discarded by real operations, so adjoining it preserves $\Dtri$. For direct sums, measuring the real orthogonal block label gives the lower bound. Convexity and invariance under real isometric embeddings give the upper bound. Finally, strong monotonicity of $M$ and its agreement with $\gtri$ on qubits give $M(\rho)\ge\sum_\mu p_\mu\gtri(\rho_\mu)$ for every admissible instrument. Taking the supremum proves minimality.
\end{proof}

The minimality statement is a specialization of the extension principle of Ref.~\cite{Gour2020}. The qubit formula applies whenever the support of $\rho$ is contained in the complex span of two real vectors. In this case a real isometry identifies the supporting subspace with a qubit. A bound of two on the complex rank alone does not ensure this property.

\begin{corollary}\label{cor:pure}
For every normalized pure state in finite dimension,
\begin{equation}
 \Dtri(\ket\psi\bra\psi)=1-|\psi^T\psi|^2=R(\psi)^2.
 \label{eq:pure}
\end{equation}
\end{corollary}
The real and imaginary parts of $\ket\psi$ span a real space of dimension at most two. Compression to that space attains the value in Eq.~\eqref{eq:pure} by the pure-qubit formula~\cite{WuPRA2021}. For the converse, set $z_\mu=\psi^TK_\mu^TK_\mu\psi$. Each conditional output is pure, so
\begin{equation*}
 \sum_\mu p_\mu\gtri(\rho_\mu)
 =1-\sum_\mu\frac{|z_\mu|^2}{p_\mu}
 \le1-\left|\sum_\mu z_\mu\right|^2,
\end{equation*}
where the inequality follows from Cauchy--Schwarz. Completeness gives $\sum_\mu z_\mu=\psi^T\psi$, which establishes the upper bound. On pure states the area thus induces the ordering already determined by $R(\psi)$.

A flagged mixture illustrates why the average over outcomes matters in higher dimensions. Consider the qutrit state
\begin{equation}
 \sigma_p=p\ket{+\ii}\bra{+\ii}+(1-p)\ket2\bra2,
 \quad \ket{+\ii}=\frac{\ket0+\ii\ket1}{\sqrt2},
 \label{eq:flag}
\end{equation}
for which $R=p$ and $\kappa=1$ when $p>0$. Assigning the value $R^2\kappa$ directly to the input would give $p^2$. A measurement of the subspace label, however, yields a qubit of unit area with probability $p$ and a real state otherwise. The direct-sum identity of Theorem~\ref{thm:D} gives $\Dtri(\sigma_p)=p$, equal to this average.
Direct comparison of two independent qutrit preparations gives a different operational quantity. At bias $t$, the weighted difference $[(1+t)\sigma_p-(1-t)\sigma_p^T]/2$ has trace norm $p+(1-p)t$ because the two states are orthogonal on the qubit block and identical on the real block. Averaging the fixed-bias success probability gives $2/3+p(1+p)/6$, strictly below $2/3+\Dtri(\sigma_p)/3$ for $0<p<1$. In direct comparison the two preparations carry independent block labels, whereas Eq.~\eqref{eq:Ddef} averages a qubit area over a single instrument outcome.

The extracted state $\ket{+\ii}$ also serves as a resource in the random real Clifford circuits studied by Magni et al.~\cite{Magni2026}, where a single copy changes the statistics of the output probabilities.

\section{Extraction bounds and exact values}\label{sec:extraction}
\subsection{Optimal extraction of squared robustness}
How much squared robustness can a real instrument extract into qubits on average? This problem admits an explicit optimal measurement. Its outcome probabilities and qubit states will also give bounds on the extractable area. Let $X$ be the unique Hermitian solution on $\supp A$ of the Sylvester equation
\begin{equation}
 AX+XA=2H,\qquad H=\ii B,
 \label{eq:SLD}
\end{equation}
with $X$ extended by zero on $\ker A$. We shall use the quantity
\begin{equation*}
 J(\rho)=\Tr(HX)=\Tr(AX^2).
\end{equation*}
In a real eigenbasis of $A$ with eigenvalues $a_j$, we have
\begin{equation}
 J(\rho)=\sum_{j,k}\frac{2|B_{jk}|^2}{a_j+a_k},
 \label{eq:Jspectral}
\end{equation}
where the sum runs over all ordered pairs $(j,k)$ with $a_j+a_k>0$. Equation~\eqref{eq:Jspectral} identifies $J$ with the Bures (minimal) quantum $\chi^2$ divergence $\chi^2_{\rm Bures}(\rho,A)$~\cite{Temme2010}. The same inverse Sylvester map enters the Bures metric~\cite{Huebner1992,Dittmann1999}, whose connection with transition probabilities goes back to Uhlmann~\cite{Uhlmann1976}. Here the divergence compares $\rho$ with its real part $A$, as in the broader class of constructions studied in Ref.~\cite{Liu2026}. For a qubit, Eq.~\eqref{eq:Jspectral} gives $J(\tau)=R(\tau)^2$. The following theorem gives $J$ an operational meaning in every finite dimension.

\begin{theorem}\label{thm:J}
For every finite-dimensional state,
\begin{equation*}
 J(\rho)=\max_{\{K_\mu\}}\sum_\mu p_\mu R(\rho_\mu)^2,
\end{equation*}
where the maximum is over the same complete real instruments with qubit outputs as in Eq.~\eqref{eq:Ddef}. It is attained by projecting onto mutually orthogonal real subspaces and encoding each outcome in a qubit. Every nonreal output comes from a two-dimensional subspace.
\end{theorem}
\begin{proof}
For each outcome of a real instrument, write $A_\mu=K_\mu A K_\mu^T$ and $H_\mu=K_\mu H K_\mu^T$. Let $X_\mu$ solve $A_\mu X_\mu+X_\mu A_\mu=2H_\mu$ on $\supp A_\mu$ and extend it by zero on the kernel. Substituting both Sylvester equations into the square of $\Delta_\mu=K_\mu X-X_\mu K_\mu$ gives
\begin{align}
 &\Tr(K_\mu^TK_\mu XAX)-\Tr(A_\mu X_\mu^2)\nonumber\\
 &\hspace{18pt}=\Tr(A\Delta_\mu^\dagger\Delta_\mu)\ge0.
 \label{eq:Jcertificate}
\end{align}
Completeness gives $\sum_\mu K_\mu^TK_\mu=\id$. Summing Eq.~\eqref{eq:Jcertificate} and using $\Tr(A_\mu X_\mu^2)=p_\mu J(\rho_\mu)$ for each outcome of positive probability yields $\sum_\mu p_\mu J(\rho_\mu)\le J(\rho)$. For qubit outputs, this is the required upper bound on the average squared robustness.

To attain the bound, conjugate Eq.~\eqref{eq:SLD}. Uniqueness on $\supp A$ gives $X^*=-X$, so $X$ is purely imaginary. The real skew-symmetric canonical form~\cite{HornJohnson2013} yields a real orthogonal matrix $O$ such that
\begin{equation}
 O^TXO=\bigoplus_jx_j\sigma_y\oplus0,\qquad x_j>0.
 \label{eq:Xblocks}
\end{equation}
Let the columns of $U_j$ span the $j$th canonical plane in the order fixed by $O$. The Kraus operator $K_j=U_j^T$ projects onto this plane and maps its basis to the qubit basis. Put $C_j=U_j^TAU_j$ and $q_j=\Tr C_j$. Restricting Eq.~\eqref{eq:SLD} to the plane and using $\{C_j,\sigma_y\}=q_j\sigma_y$ gives
\begin{equation*}
 U_j^THU_j=\frac{q_jx_j}{2}\sigma_y,\qquad R(\rho_j)=x_j.
\end{equation*}
One-dimensional projections along a real orthonormal basis of $\ker X$ complete the measurement. These outcomes can be encoded as real qubit states and contribute zero robustness. The mean squared robustness is therefore
\begin{equation*}
 \sum_jq_jR(\rho_j)^2=\sum_jq_jx_j^2=\Tr(AX^2)=J(\rho).
\end{equation*}
\end{proof}

The same measurement gives an explicit area yield. The $j$th plane occurs with probability $q_j$, has robustness $x_j$, and has maximum conditional robustness $\kappa_j=q_jx_j/(2\sqrt{\det C_j})$. Weighting its area by $q_j$ gives the lower bound
\begin{equation}
 L_X(\rho)=\sum_j\frac{q_j^2x_j^3}{2\sqrt{\det C_j}}
 \le\Dtri(\rho).
 \label{eq:LX}
\end{equation}
Degenerate canonical blocks allow a choice of planes in Eq.~\eqref{eq:Xblocks}. Each choice gives an attainable value in Eq.~\eqref{eq:LX}. This protocol attains the optimal area for every qubit and for pure inputs in any dimension. For a general mixed input, Theorem~\ref{thm:J} establishes optimality for squared robustness alone.

\subsection{Area bounds and local estimation}
\begin{figure}[t]
 \centering
 \includegraphics[width=\columnwidth]{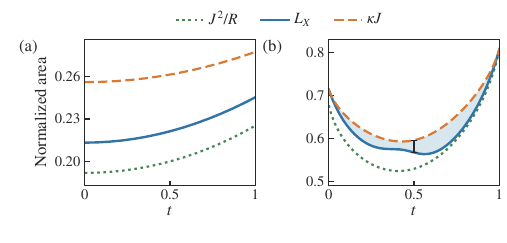}
 \caption{Bounds on the extractable area. (a) For $A=\id_6/6$ and block strengths $(0.8,0.4+0.2t,0.4-0.2t)$, the projective yield $L_X$ equals the exact area $(80+12t^2)/375$. (b) For the noncommuting family in Eq.~\eqref{eq:boundsfamily}, the shaded band between $L_X$ and $\kappa J$ contains the optimum. The marked interval at $t=1/2$ extends from $L_X\simeq0.5672$ to $\kappa J\simeq0.5957$. The dotted curve gives the lower bound $J^2/R$ in both panels.}
 \label{fig:bounds}
\end{figure}
\begin{theorem}\label{thm:bounds}
If $R(\rho)>0$, then
\begin{equation}
 \frac{J(\rho)^2}{R(\rho)}\le\Dtri(\rho)\le\kappa(\rho)J(\rho).
 \label{eq:bounds}
\end{equation}
Moreover,
\begin{equation}
 R^2\le J\le\kappa R,\qquad R^3\le\Dtri\le\kappa^2R\le R.
 \label{eq:chain}
\end{equation}
All these quantities vanish on real states.
\end{theorem}
\begin{proof}
For the projective measurement of Theorem~\ref{thm:J}, write $m_k=\sum_jq_jx_j^k$. We have $m_2=J$ and, by strong monotonicity of robustness, $m_1\le R$. Each output has $\kappa_j\ge R_j=x_j$, so $L_X\ge m_3$. Cauchy--Schwarz now gives $J^2=m_2^2\le m_1m_3\le R L_X$, proving the lower bound. For any real instrument with qubit outputs, branchwise monotonicity of $\kappa$ and Theorem~\ref{thm:J} give
\begin{equation*}
 \sum_\mu p_\mu R_\mu^2\kappa_\mu
 \le\kappa\sum_\mu p_\mu R_\mu^2\le\kappa J,
\end{equation*}
and maximizing gives the upper bound in Eq.~\eqref{eq:bounds}. The Bures $\chi^2$ inequality~\cite{Temme2010} gives $R^2\le J$ (see also Appendix~\ref{app:bounds}). By positivity of the inverse Sylvester map, Eq.~\eqref{eq:order} implies $\norm X_\infty\le\kappa$. Thus $J=\Tr(HX)\le\kappa\norm H_1=\kappa R$. Substitution into Eq.~\eqref{eq:bounds} gives Eq.~\eqref{eq:chain}.
\end{proof}

The upper bound is exact for every qubit. The bounds coincide for qubits on the $y$ axis ($r_x=r_z=0$) and for the flagged family in Eq.~\eqref{eq:flag}.

Figure~\ref{fig:bounds} illustrates what can be learned when the bounds differ. In panel (a), neither $R$ nor $\kappa$ varies along the commuting family, yet redistributing imaginarity between its planes changes the exact area given by Proposition~\ref{prop:commuting}. For the block strengths shown in panel (a), $R=8/15$, $\kappa=4/5$, and $J=(24+2t^2)/75$. Panel (b) considers the four-dimensional noncommuting family
\begin{align}
 \eta_t&=(1-\epsilon)\bigl[t\ket u\bra u+(1-t)\ket v\bra v\bigr]
       +\tfrac\epsilon4\id_4,\nonumber\\
 \ket u&=\frac{\ket0+\ii\ket1}{\sqrt2},\qquad
 \ket v=\frac{\ket0+\ket2+\ii\ket3}{\sqrt3},
 \label{eq:boundsfamily}
\end{align}
with $\epsilon=0.10$ and $0\le t\le1$. Here the projective yield $L_X$ tightens the scalar lower bound $J^2/R$. We evaluated $L_X$ by solving Eq.~\eqref{eq:SLD} in a real eigenbasis of $A$ and reducing $\operatorname{Im}X$ to real Schur form. The interval from $L_X$ to $\kappa J$ locates the optimum using this instrument.

The same projectors have an interpretation in local parameter estimation~\cite{Helstrom1976,Holevo1982}. Suppose two sources prepare the known states $\rho$ and $\rho^T$. A parameter $\theta$ controls how often each source is selected, giving the family
\begin{equation}
 \rho_\theta=\frac{1+\theta}{2}\rho+\frac{1-\theta}{2}\rho^T
 =A+\ii\theta B,\qquad |\theta|\le1.
 \label{eq:model}
\end{equation}
This preparation requires only classical mixing of the two known states. At $\theta=0$, the state is $A$ and its derivative is $H$. Equation~\eqref{eq:SLD} thus identifies $X$ as the symmetric logarithmic derivative at the midpoint. The corresponding quantum Fisher information is~\cite{Braunstein1994,Paris2009}
\begin{equation*}
 F_Q[\rho_\theta]_{\theta=0}=J(\rho).
\end{equation*}
The argument $\rho$ in $J(\rho)$ specifies the endpoint that fixes the direction $H$ of the model. Although the midpoint is real, the local information there depends on this direction~\cite{LiuQFI2020}.

\begin{figure}[t]
 \centering
 \includegraphics[width=\columnwidth]{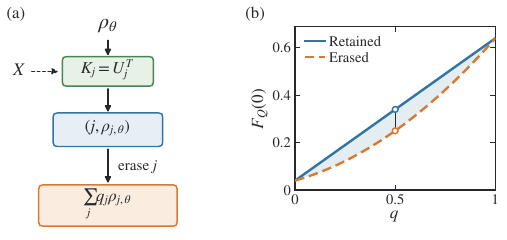}
 \caption{Projective extraction and the outcome record. (a) The dashed arrow indicates that $X$ fixes the measurement planes. The real instrument maps $\rho_\theta$ to a qubit with classical label $j$. Erasing $j$ averages the qubit states in a common basis. (b) Midpoint quantum Fisher information for $\omega_q=q\tau_{0.8}\oplus(1-q)\tau_{0.2}$ with general quantum readout. Retaining the label gives $0.04+0.60q$, and erasing it gives $(0.2+0.6q)^2$. The marked points at $q=1/2$ give the largest loss, $0.09$.}
 \label{fig:protocol}
\end{figure}

The projectors in Fig.~\ref{fig:protocol}(a) depend only on the known pair, and each plane has probability $q_j$ for every $\theta$. Keeping the outcome label therefore gives midpoint quantum Fisher information $\sum_jq_jx_j^2=J$ in the qubit outputs. This value assumes general quantum measurements at readout. For a real POVM effect $M$, $\Tr(MB)=0$ and the outcome probability is independent of $\theta$.

To see the effect of erasing the record, consider $\omega_q=q\tau_a\oplus(1-q)\tau_b$, with $\tau_c=(\id+c\sigma_y)/2$, $a=0.8$, and $b=0.2$. The block weight $q$ is known, and $\theta$ is the parameter to be estimated. Keeping the two outcomes gives midpoint information $F_Q=qa^2+(1-q)b^2$. Mapping both blocks to the same qubit basis and erasing the label gives $F_Q=[qa+(1-q)b]^2$. Their difference, $q(1-q)(a-b)^2$, is the variance of the two block strengths [Fig.~\ref{fig:protocol}(b)].

Lu et al.~\cite{Lu2026} studied imaginarity through a channel that encodes an unknown $p\in[0,1]$, with $0<m\le1/(d+1)$ fixed. Their encoding gives a direct comparison with Eq.~\eqref{eq:model}. For the depolarized state $\sigma_m=m\rho+(1-m)\id_d/d$ and its real part $A_m=mA+(1-m)\id_d/d$, their channel output takes the form
\begin{align*}
 \Lambda_{m,p}(\rho)
 &=A_m+(p-1)mH\\
 &=\frac p2\sigma_m+\frac{2-p}{2}\sigma_m^T.
\end{align*}
This is the part $-1\le\theta\le0$ of the conjugate-mixture model with endpoint $\sigma_m$ and coordinate $\theta=p-1$. The point $p=1$ corresponds to its real midpoint. Since $d\theta/dp=1$, the quantum Fisher information there is
\begin{equation}
 F_{m,1}(\rho)=J(\sigma_m).
 \label{eq:LuEndpoint}
\end{equation}
The one-sided derivative at $p=1$ is regular because $A_m>0$. For a qubit the result is $m^2J(\rho)=m^2R(\rho)^2$. In higher dimensions, depolarization generally also changes the inverse Sylvester map in Eq.~\eqref{eq:SLD}.

Lu et al. established monotonicity of their estimation quantity under real channels covariant with their encoding, equivalently under real unital channels. Theorem~\ref{thm:J} gives the endpoint value in Eq.~\eqref{eq:LuEndpoint} a further interpretation. Applied to $\sigma_m$, the projective measurement attains $J(\sigma_m)$ as the maximum mean squared robustness over all real instruments with qubit outputs. For the area, the connection to estimation is through the bounds in Eq.~\eqref{eq:bounds}.

\subsection{Exactly solvable families}
\begin{figure}[t]
 \centering
 \includegraphics[width=\columnwidth]{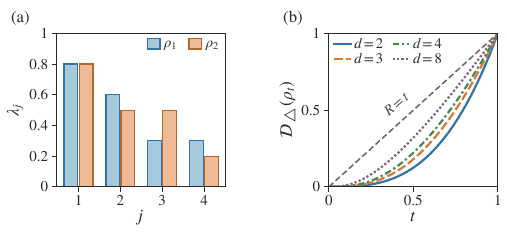}
 \caption{Exactly solvable families. (a) Paired bars give the strengths of equally weighted qubit blocks in $\rho_1$ and $\rho_2$. Their mean, maximum, and mean square agree, but their mean cubes give extractable areas differing by $3/1000$ [Eq.~\eqref{eq:eightdim}]. (b) Exact areas for a maximally imaginary pure state of weight $t$ mixed with white noise [Eq.~\eqref{eq:depolmax}]. The dashed diagonal is the upper bound $R=t$.}
 \label{fig:two}
\end{figure}
Orthogonal block structure makes the extraction problem particularly simple. Suppose that real projectors resolve the block label and that each nonreal block is a qubit. Direct-sum additivity then gives
\begin{equation}
 \Dtri\!\left(\bigoplus_jq_j\tau_j\right)=\sum_jq_j\gtri(\tau_j)
 \label{eq:directsum}
\end{equation}
where real one-dimensional blocks contribute zero. Two families admit such a decomposition.

\begin{proposition}\label{prop:commuting}
If $[A,B]=0$, then
\begin{equation}
 \Dtri(\rho)=\Tr(|B|^3A^{-2}).
 \label{eq:commuting}
\end{equation}
In particular, for $A=\id_d/d$,
\begin{equation*}
 \Dtri(\rho)=d^2\Tr|B|^3.
\end{equation*}
\end{proposition}
\begin{proof}
Since $A$ and $B$ commute, $B$ preserves every real eigenspace of $A$. Within each eigenspace a real orthogonal change of basis puts $B$ in its skew-symmetric canonical form. On each nonreal block we have $A_j=a_j\id_2$ and $B_j=b_j\bigl(\begin{smallmatrix}0&1\\-1&0\end{smallmatrix}\bigr)$. The block has weight $2a_j$ and its normalized state has $R_j=\kappa_j=|b_j|/a_j$. Equation~\eqref{eq:directsum} therefore gives $\sum_j2|b_j|^3/a_j^2$, the trace in Eq.~\eqref{eq:commuting}.
\end{proof}

Figure~\ref{fig:two}(a) illustrates which features of the block distribution enter the area for a commuting state. Consider four equally weighted qubits $\tau_\lambda=(\id_2+\lambda\sigma_y)/2$. The state $\rho(\bm\lambda)=\bigoplus_{j=1}^4\tfrac14\tau_{\lambda_j}$, with $0\le\lambda_j\le1$, has real part $A=\id_8/8$. Its four extraction quantities are
\begin{align}
 R=\tfrac14\sum_j\lambda_j,\quad
 &\kappa=\max_j\lambda_j,\nonumber\\
 J=\tfrac14\sum_j\lambda_j^2,\quad
 &\Dtri=\tfrac14\sum_j\lambda_j^3.
 \label{eq:moments}
\end{align}
Thus $R$ is the mean strength of a block selected with its physical probability, and $\kappa$ is the strength of the strongest block. The optimal quadratic yield $J$ is the second moment, while the extractable area is the third. For example, the block strengths $\bm\lambda^{(1)}=(0.8,0.6,0.3,0.3)$ and $\bm\lambda^{(2)}=(0.8,0.5,0.5,0.2)$ give the same $R=1/2$, $\kappa=4/5$, and $J=59/200$. Their areas are
\begin{equation}
 \Dtri(\rho_1)=\frac{391}{2000},\qquad
 \Dtri(\rho_2)=\frac{77}{400}.
 \label{eq:eightdim}
\end{equation}
Moving strength among the three weaker blocks increases the area of $\rho_1$ over that of $\rho_2$ by $3/1000$ [Fig.~\ref{fig:two}(a)]. Thus no deterministic real operation takes $\rho_2$ to $\rho_1$, although $R$, $\kappa$, and $J$ coincide. Since both real parts are maximally mixed, Eq.~\eqref{eq:LuEndpoint} gives the same endpoint value $F_{m,1}=m^2J$ for every allowed $m$. Equality of $F_{m,p}$ away from $p=1$ does not follow.

\begin{proposition}\label{prop:depol}
Let $d\ge2$ and $\rho_t=t\ket\psi\bra\psi+(1-t)\id_d/d$, with $0\le t\le1$. Write
\begin{equation*}
 r=\sqrt{1-|\psi^T\psi|^2},\qquad \nu=(1-t)/d,\qquad w=t+2\nu.
\end{equation*}
Then
\begin{equation}
 \Dtri(\rho_t)=\frac{t^3r^3}{w\sqrt{t^2r^2+4\nu t+4\nu^2}}.
 \label{eq:depol}
\end{equation}
At $r=0$ the formula is understood by continuity and gives zero.
\end{proposition}
\begin{proof}
Choose a global phase and a real orthogonal basis so that $\ket\psi=\cos\alpha\ket0+\ii\sin\alpha\ket1$. Only the plane spanned by $\ket0$ and $\ket1$ can then carry an imaginary part. The block on this plane has weight $w$, and the remaining $d-2$ blocks are real and one dimensional. Before normalization, the magnitude of the imaginary off-diagonal entry in the plane is $tr/2$, and the determinant of the real part is $\nu^2+\nu t+t^2r^2/4$. Evaluating the normalized qubit area in Eq.~\eqref{eq:g} and multiplying by $w$ as in Eq.~\eqref{eq:directsum} gives Eq.~\eqref{eq:depol}.
\end{proof}

For a maximally imaginary pure component, $r=1$ and Eq.~\eqref{eq:depol} becomes
\begin{equation}
 \Dtri(\rho_t)=\frac{t^3}{[t+2(1-t)/d]^2}.
 \label{eq:depolmax}
\end{equation}
Figure~\ref{fig:two}(b) shows the dependence on $d$ at fixed pure-state weight $t$. The two-dimensional block receives only $2(1-t)/d$ of the white noise. As $d$ increases, a larger fraction of the noise falls in the orthogonal real subspace, which contributes zero to the average area. The curves therefore compare different preparations.

\subsection{Finite-outcome optimization}
For a general state, the number of outcomes needed for optimal extraction can be bounded in terms of $s=\rank A$. To see how the bound arises, restrict the input space to the real support of $A$. Each qubit outcome contributes to the average through its effect $E_\mu=K_\mu^TK_\mu$, a real positive semidefinite matrix of rank at most two. For such an effect $E$, define
\begin{align*}
 p(E)&=\Tr(AE),\qquad z(E)=-\tfrac12\Tr(BEBE),\\
 \delta(E)&=\tfrac12\{p(E)^2-\Tr(AEAE)\},\\
 f_\rho(E)&=\frac{4z(E)^{3/2}}{p(E)\sqrt{\delta(E)}}.
\end{align*}
Both $z(E)$ and $\delta(E)$ are nonnegative. At degenerate outcomes $f_\rho(E)=0$ by continuity.

\begin{proposition}\label{prop:finite}
There is an optimal instrument with at most $s(s+1)/2$ outcomes of nonzero probability. Let $\mathcal E_2$ denote the set of real positive semidefinite $s\times s$ matrices of rank at most two. The optimal area is
\begin{equation}
 \Dtri(\rho)=\max_{\{E_\mu\}}\sum_\mu f_\rho(E_\mu),
 \label{eq:effectopt}
\end{equation}
where the maximum is over finite families of effects $E_\mu\in\mathcal E_2$ satisfying $\sum_\mu E_\mu=\id_s$. Equivalently, the extractable area is the attained minimum
\begin{equation}
 \Dtri(\rho)=\min_{Y=Y^T}\left\{\Tr Y:
 \Tr(YE)\ge f_\rho(E)\ \forall E\in\mathcal E_2\right\}.
 \label{eq:dual}
\end{equation}
\end{proposition}
Appendix~\ref{app:finite} proves the proposition and expresses the dual constraint as a scalar inequality on each real two-dimensional plane. An admissible set of effects in Eq.~\eqref{eq:effectopt} gives a lower bound, and a matrix $Y$ satisfying Eq.~\eqref{eq:dual} gives an upper bound. The optimization over effects is finite dimensional and nonconvex. To certify the dual bound, the inequality must hold on every plane, including those outside any finite sample.

\section{Discussion}\label{sec:discussion}
Equal qubit areas give equal mean success probabilities in Corollary~\ref{cor:comparison}, even for states that are incomparable under deterministic real operations. For example, the pairs $(R,\kappa)=(1/2,1/2)$ and $(1/(2\sqrt{2}),1)$ both give $\gtri=1/8$. The first state has greater initial robustness, while only the second can yield a maximally imaginary qubit on a successful filter outcome. A deterministic conversion in either direction would increase one of the two parameters in Eq.~\eqref{eq:conversion}.

The eight-dimensional states in Eq.~\eqref{eq:eightdim} show how the area can rule out a conversion even when $R$, $\kappa$, and $J$ agree. Postselecting the strongest block gives the same qubit with the same probability in both preparations. Redistributing strength among the other three blocks preserves both the mean and the mean square. On each block in this family, $R_j=\kappa_j=\lambda_j$, so a block of probability $q_j$ contributes $q_j\lambda_j^3$ to the area. The third moment thus weights the stronger blocks more heavily than the quadratic yield does. This redistribution gives $\rho_1$ an area larger than that of $\rho_2$ by $3/1000$, ruling out the deterministic real conversion $\rho_2\to\rho_1$.

When does the projection in Theorem~\ref{thm:J} also attain the optimal area? For an outcome of probability $q_j$, let $x_j$ be the output robustness and $\kappa_j$ the largest robustness attainable by further filtering that output. The difference between the upper bound and the area attained by the projection is $\kappa J-L_X=\sum_j q_jx_j^2(\kappa-\kappa_j)$. Each term is nonnegative. If every outcome with $q_jx_j^2>0$ retains the input value of $\kappa$, the difference vanishes and the projection is optimal for area. When the difference is positive, it bounds the gain that any alternative instrument could achieve. For the eight-dimensional pair the gain is zero. Reading the block label attains both exact areas, although the common upper bound $\kappa J=59/250$ exceeds both values.

The plane inequality in Appendix~\ref{app:finite} offers a way to sharpen the upper bound for general mixed states. One can search for a dual matrix whose trace is close to the area already attained by a candidate instrument. Its trace then certifies an upper bound if the inequality holds on every plane, including the unsampled ones.

Corollary~\ref{cor:comparison} gives a task interpretation for the qubit area. Its extension averages over extracted qubits, and the qutrit example in Eq.~\eqref{eq:flag} shows why direct comparison of higher-dimensional inputs gives a different answer. It remains open whether a task on the original inputs can quantify the full extractable area in arbitrary dimension. Tasks involving hiding and masking~\cite{Zhu2021}, broadcasting~\cite{ZhangBroadcast2024}, conversion into entanglement~\cite{Sun2025}, or entanglement concentration with a nonreal measurement~\cite{Biswas2026} provide settings in which to ask this question. For channel imaginarity~\cite{ChenLei2025}, an extraction task would also have to specify which probe states may be supplied to the channel.

\section*{AI USAGE}
ChatGPT was used solely for polishing the text and checking some mathematical derivations; all intellectual contributions---the conceptual framework, main ideas, and proofs---were completed independently by the authors without the assistance of artificial intelligence.

\section*{acknowledgments}
This work was supported by the National Natural Science Foundation of China under Grant No.~12301582; the Jiangxi Provincial Natural Science Foundation under Grant No.~20252BAC200156; the Early-Career Young Scientists and Technologists Project of Jiangxi Province under Grant No.~20244BCE52197; the Doctoral Research Startup Fund Project of East China University of Technology under Grant No.~DHBK2024029.

\appendix

\section{Convexity and strong monotonicity for qubits}
\label{app:qubit}

Write $y=r_y$ and $\boldsymbol v=(r_x,r_z)$. The denominator in Eq.~\eqref{eq:g} is $a(\boldsymbol v)=\sqrt{1-\norm{\boldsymbol v}^2}$, a concave function on the unit disk. Consider the function
\begin{equation*}
 f(y,a)=\frac{|y|^3}{a},\qquad a>0,
\end{equation*}
whose Hessian for $y>0$ has diagonal entries $f_{yy}=6y/a$ and $f_{aa}=2y^3/a^3$. The mixed derivative is $f_{ya}=-3y^2/a^2$, giving a Hessian determinant of $3y^4/a^4$. Thus the Hessian is positive semidefinite. Evenness in $y$ gives the same conclusion for $y<0$, and the second derivatives extend continuously to $y=0$. Hence $f$ is jointly convex. It is also nonincreasing in $a$. For a mixture of Bloch vectors, concavity of $a$ therefore allows us to replace the denominator by the average of the component denominators to obtain an upper bound. Joint convexity of $f$ bounds this in turn by the average of the component areas. This proves convexity wherever $a>0$. Since $|y|\le a$ on the Bloch ball, $0\le f(y,a)\le y^2\to0$ as $a\to0$, so convexity extends to the boundary.

We next consider a qubit state $\tau$ with real part $A$ of full rank. Let $K$ be a real $2\times 2$ Kraus operator, $E=K^TK$, $p=\Tr(EA)>0$, and $d_K=|\det K|$. If $K$ is invertible, each real antisymmetric $2\times2$ matrix is multiplied by $\det K$ under congruence by $K$. Since $a=2\sqrt{\det A}$, the normalized output $\tau'$ obeys
\begin{equation*}
 R'=\frac{d_KR}{p},\qquad
 a'=\frac{d_Ka}{p},\qquad
 \kappa'=\kappa.
\end{equation*}
It follows that
\begin{equation}
 p\gtri(\tau')=
 \frac{\det E}{\Tr(EA)}\gtri(\tau).
 \label{eq:gbranch}
\end{equation}
A rank-one $K$ gives a real output, so both sides vanish in this case. To sum the branch contributions, choose a real orthonormal basis in which $A=\operatorname{diag}(t,1-t)$ and write $E=\bigl(\begin{smallmatrix}e&f\\f&h\end{smallmatrix}\bigr)$. Then
\begin{align*}
 &\Tr(EA)\Tr[E(\id-A)]-\det E\\
 &\qquad=t(1-t)(e-h)^2+f^2\ge0.
\end{align*}
Dividing by $\Tr(EA)$ and summing over the outcomes of a complete instrument gives
\begin{equation*}
 \sum_\mu\frac{\det E_\mu}{\Tr(E_\mu A)}
 \le\sum_\mu\Tr[E_\mu(\id-A)]=1,
\end{equation*}
which, together with Eq.~\eqref{eq:gbranch}, proves Eq.~\eqref{eq:gstrong}. We omit zero-probability outcomes. If $A$ is singular, the input qubit is real and every output has zero area.

\section{Maximum conditional robustness}
\label{app:support}

If $v$ is a real vector in $\ker A$, then $v^\dagger\rho v=0$. Positivity gives $\rho v=0$ and hence $Bv=0$. As $\ker A$ has a real orthonormal basis, $\supp B\subseteq\supp A$. On this support, the two inequalities $\rho=A+H\ge0$ and $\rho^T=A-H\ge0$ give
\begin{equation*}
 -\id\le S\le\id,\qquad S=A^{-\frac{1}{2}}HA^{-\frac{1}{2}},
\end{equation*}
so $\kappa=\norm S_\infty\le1$. The real skew-symmetric matrix $S/\ii$ has a real orthogonal decomposition into two-dimensional blocks on its support and a zero block on its kernel.

To bound the robustness, take the trace of $H\le\kappa A$ against the positive spectral projector of $H$, and of $-H\le\kappa A$ against its negative spectral projector. Adding the inequalities yields $\norm H_1\le\kappa\Tr A=\kappa$, as claimed in Eq.~\eqref{eq:order}. A real completely positive branch preserves the inequalities $-\kappa A\le H\le\kappa A$. Normalizing the output by its probability preserves them as well. The output thus has conditional parameter at most $\kappa$ and robustness at most $\kappa$.

For $\kappa=0$, every output is real and the bound is attained. For $\kappa>0$, the support has dimension $s=\rank A\ge2$. Choose a real isometry $U\in\RR^{s\times2}$ onto a block of $S$ at which its norm is attained, so that $U^TSU=\pm\kappa\sigma_y$. For sufficiently small $c>0$, the operator
\begin{equation*}
 K=cU^TA^{-\frac{1}{2}},\qquad K^TK\le\id_s,
\end{equation*}
defines a success branch. The unnormalized output is $c^2\id_2\pm c^2\kappa\sigma_y$, with probability $2c^2>0$. Its normalized robustness is exactly $\kappa$. A decomposition of $\id_s-K^TK$ into real rank-one effects supplies the remaining qubit outcomes. On the original input space, any additional outcomes needed to cover $\ker A$ have zero probability.

\section{Bounds on \texorpdfstring{$J$}{J} and midpoint quantum Fisher information}
\label{app:bounds}

On $\supp A$, the solution of Eq.~\eqref{eq:SLD} can be written as
\begin{equation*}
 X=2\int_0^\infty e^{-tA}H e^{-tA}\,dt.
\end{equation*}
The linear map on the right preserves order and takes $A$ to the identity. Applying it to $-\kappa A\le H\le\kappa A$ gives $-\kappa\id\le X\le\kappa\id$. Thus $J=\Tr(HX)\le\kappa\norm H_1=\kappa R$.

For the other bound, take $S_H=\operatorname{sgn}(H)$. Equation~\eqref{eq:SLD} gives
\begin{equation*}
 R=\Tr(S_HH)=\operatorname{Re}\Tr(AS_HX).
\end{equation*}
By the Hilbert--Schmidt Cauchy--Schwarz inequality and $S_H^2\le\id$,
\begin{equation*}
 R^2\le\Tr(AS_H^2)\Tr(AX^2)\le J.
\end{equation*}
When $R=0$, both $H$ and $X$ vanish and both bounds hold with equality.

The branch identity in Eq.~\eqref{eq:Jcertificate} applies to rectangular Kraus operators with outputs of any finite dimension. Indeed, $H_\mu$ is supported on $\supp A_\mu$, so the solution of the output Sylvester equation on that support also solves it on the full space when extended by zero. Expanding $\Delta_\mu^\dagger\Delta_\mu$ under the trace and using the input and output Sylvester equations for the cross terms gives Eq.~\eqref{eq:Jcertificate}. For the upper bound in Eq.~\eqref{eq:bounds}, consider the dual matrix
\begin{equation*}
 Y_* = \kappa XAX,\qquad \Tr Y_* = \kappa J.
\end{equation*}
The matrix $XAX$ is real symmetric and positive semidefinite. For each qubit extraction branch, $J(\rho_\mu)=R(\rho_\mu)^2$ and $\kappa(\rho_\mu)\le\kappa$. The branch identity then gives $f_\rho(K^TK)\le\Tr(Y_*K^TK)$.

For the local estimation model, keep the outcome label of the projective instrument of Theorem~\ref{thm:J}. Its output is
\begin{equation*}
 \Omega_\theta=
 \bigoplus_j q_j\tau_{j,\theta}\ \oplus\ \Omega_{\ker}.
\end{equation*}
Here $j$ labels the nonzero two-dimensional blocks of $X$, with real isometries $U_j$. Their probabilities $q_j$ are independent of $\theta$ since $\Tr(U_j^TBU_j)=0$. On $\ker X$, the compression of $H=(AX+XA)/2$ vanishes. Hence the corresponding unnormalized output $\Omega_{\ker}$ is independent of $\theta$ and can be refined into real qubit outcomes. At $\theta=0$, the symmetric logarithmic derivative of the $j$th normalized block is $x_j\sigma_y$. Since the blocks carry distinct classical labels, summing their contributions gives
\begin{align*}
 F_Q[\Omega_\theta]_{\theta=0}
 &=\sum_j q_j\Tr[\tau_{j,0}(x_j\sigma_y)^2]\\
 &=\sum_j q_jx_j^2=J.
\end{align*}
The projective protocol thus preserves the quantum Fisher information at the real midpoint of this one-parameter model.

\section{Finite attainment and duality}
\label{app:finite}

We restrict the input to the real support of $A$, where $A>0$. For $s=\rank A=1$ the state is real, one outcome suffices, and the dual value is attained by the zero matrix. Let $s\ge2$ and $n=s(s+1)/2$.

For a branch $K\in\RR^{2\times s}$, write $C=KAK^T$, $E=K^TK$, and $KBK^T=\bigl(\begin{smallmatrix}0&\beta\\-\beta&0\end{smallmatrix}\bigr)$. Substitution into the definitions of $p$, $z$, and $\delta$ gives
\begin{equation*}
 p=\Tr C=\Tr(AE),\qquad z=\beta^2,\qquad \delta=\det C.
\end{equation*}
For $\delta>0$, the conditional qubit has $R'=2|\beta|/p$ and $\kappa'=|\beta|/\sqrt\delta$. Its contribution is therefore $p\gtri(\rho')=f_\rho(E)$. If $\delta=0$, positivity forces $\beta=0$, and the contribution vanishes. Conversely, each real positive semidefinite $E$ of rank at most two admits a factorization $E=K^TK$ with $K\in\RR^{2\times s}$. Optimizing over complete instruments is therefore equivalent to optimizing over their effects.

The function $f_\rho$ is positively homogeneous and satisfies $0\le f_\rho(E)\le\Tr(AE)=p$. To see continuity at degenerate effects, note that positivity of the unnormalized qubit gives $z\le\delta$. If $\delta\to0$ with $p$ bounded away from zero, then $f_\rho(E)\le4\delta/p\to0$. If $p\to0$, the bound $f_\rho(E)\le p$ suffices. The graph over trace-one effects,
\begin{equation*}
 \mathcal K=
 \{(e,f_\rho(e)):e\ge0,\ \rank e\le2,\ \Tr e=1\}
\end{equation*}
is therefore compact. Write each nonzero effect as $E_\mu=s w_\mu e_\mu$, with $\Tr e_\mu=1$. Completeness is equivalent to $\sum_\mu w_\mu e_\mu=\id_s/s$, which also gives $\sum_\mu w_\mu=1$. By homogeneity, the second coordinate of this convex combination in $\conv\mathcal K$ is the instrument's average area divided by $s$. The largest such coordinate is attained because the convex hull is compact.

For each real density matrix $Q$, define
\begin{equation*}
 F(Q)=\max_{\{(w_\mu,e_\mu)\}}\sum_\mu w_\mu f_\rho(e_\mu),
\end{equation*}
where the maximum runs over all finite decompositions $Q=\sum_\mu w_\mu e_\mu$ with $w_\mu\ge0$ and $\sum_\mu w_\mu=1$. Each $e_\mu$ is a real density matrix of rank at most two. The maximum exists by the same compactness argument. Every real density matrix is a convex combination of real rank-one projectors, so $F$ is finite on the full set of real density matrices. Mixing two decompositions shows that $F$ is concave. The point $Q_0=\id_s/s$ lies in the relative interior of its domain, where a finite concave function admits a supergradient~\cite{Rockafellar1970}. Thus some real symmetric $G$ satisfies
\begin{equation*}
 F(Q)\le F(Q_0)+\Tr[G(Q-Q_0)].
\end{equation*}
Set $Y=G+[F(Q_0)-\Tr(GQ_0)]\id_s$. For any normalized effect $e$ of rank at most two, $f_\rho(e)\le F(e)\le\Tr(Ye)$. Positive homogeneity extends the constraint to every $E\in\mathcal E_2$. Moreover, $\Tr Y=sF(Q_0)=\Dtri(\rho)$. Weak duality now gives Eq.~\eqref{eq:dual} and attainment of the minimum. Applying the constraints to rank-one projectors also gives $Y\ge0$.

Each positive-weight effect $e$ in an optimal decomposition must satisfy $f_\rho(e)=\Tr(Ye)$, since all the dual inequalities sum to equality. The corresponding points $(e,f_\rho(e))$ lie on the graph of a linear function over real symmetric matrices of trace one, an affine space of dimension $n-1$. Carath\'eodory's theorem gives an optimal decomposition with at most $n$ effects. This bounds the number of outcomes of positive probability. Extending the instrument to $\ker A$ adds only outcomes of zero probability.

Finally, we reduce all rank-two dual constraints on a fixed plane to a single inequality. For a real isometry $U\in\RR^{s\times2}$, write $E=UMU^T$ and set
\begin{equation*}
 C=U^TAU,\qquad Z=U^TYU,\qquad
 b=|(U^TBU)_{12}|.
\end{equation*}
For $C>0$ and $Y\ge0$, the dual constraints on this plane are equivalent to
\begin{equation*}
 \Tr(\operatorname{adj}C\,Z)+2\sqrt{\det C\det Z}
 \ge\frac{4b^3}{\sqrt{\det C}}.
\end{equation*}
When $M>0$, multiply the dual inequality by $\Tr(CM)/\det M$. Its left-hand side becomes $\Tr(CM)\Tr(ZM)/\det M$ and its right-hand side is $4b^3/\sqrt{\det C}$. We use the identity
\begin{align*}
 &\inf_{M>0}\frac{\Tr(CM)\Tr(ZM)}{\det M}\\
 &\qquad=\Tr(\operatorname{adj}C\,Z)
       +2\sqrt{\det C\det Z}.
\end{align*}
To prove the identity, set $N=C^{1/2}MC^{1/2}$ and $W=C^{-1/2}ZC^{-1/2}$. In an eigenbasis of $W$, replacing $N$ by its diagonal part preserves $\Tr N$ and $\Tr(WN)$ and can only increase $\det N$. It therefore suffices to minimize over the ratio of the two positive diagonal entries. This gives $\det C\,[\Tr W+2\sqrt{\det W}]$, as required. For singular $Z$, the same value follows by taking the limit of the diagonal ratio, with the case $Z=0$ immediate. If a plane with singular $C$ is considered before restriction to $\supp A$, positivity of the compressed state forces $b=0$. Every effect on that plane then contributes zero area.

\par
\interlinepenalty=10000
\makeatletter
\let\auto@bib@innerbib\@empty
\makeatother


\begin{thebibliography}{99}
\bibitem{WuPRA2021} K.-D. Wu, T. V. Kondra, S. Rana, C. M. Scandolo, G.-Y. Xiang, C.-F. Li, G.-C. Guo, and A. Streltsov, Resource theory of imaginarity: Quantification and state conversion, Phys. Rev. A \textbf{103}, 032401 (2021). 

\bibitem{GuoCoherence2026} M.-L. Guo, N. Guan, B. Li, B. Hu, and S.-M. Fei, Imaginarity as a resource within quantum coherence: Geometric decomposition and operational conversion, Phys. Rev. A \textbf{114}, 012435 (2026). 

\bibitem{Caves2001} C. M. Caves, C. A. Fuchs, and P. Rungta, Entanglement of formation of an arbitrary state of two rebits, Found. Phys. Lett. \textbf{14}, 199 (2001). 

\bibitem{McKague2009} M. McKague, M. Mosca, and N. Gisin, Simulating quantum systems using real Hilbert spaces, Phys. Rev. Lett. \textbf{102}, 020505 (2009). 

\bibitem{Renou2021} M.-O. Renou, D. Trillo, M. Weilenmann, T. P. Le, A. Tavakoli, N. Gisin, A. Ac{\'i}n, and M. Navascu{\'e}s, Quantum theory based on real numbers can be experimentally falsified, Nature \textbf{600}, 625 (2021). 

\bibitem{LiReal2022} Z.-D. Li, Y.-L. Mao, M. Weilenmann, A. Tavakoli, H. Chen, L. Feng, S.-J. Yang, M.-O. Renou, D. Trillo, T. P. Le, N. Gisin, A. Ac{\'i}n, M. Navascu{\'e}s, Z. Wang, and J. Fan, Testing real quantum theory in an optical quantum network, Phys. Rev. Lett. \textbf{128}, 040402 (2022). 

\bibitem{ChenReal2022} M.-C. Chen, C. Wang, F.-M. Liu, J.-W. Wang, C. Ying, Z.-X. Shang, Y. Wu, M. Gong, H. Deng, F.-T. Liang, Q. Zhang, C.-Z. Peng, X. Zhu, A. Cabello, C.-Y. Lu, and J.-W. Pan, Ruling out real-valued standard formalism of quantum theory, Phys. Rev. Lett. \textbf{128}, 040403 (2022). 

\bibitem{Hickey2018} A. Hickey and G. Gour, Quantifying the imaginarity of quantum mechanics, J. Phys. A: Math. Theor. \textbf{51}, 414009 (2018). 

\bibitem{Chitambar2019} E. Chitambar and G. Gour, Quantum resource theories, Rev. Mod. Phys. \textbf{91}, 025001 (2019). 

\bibitem{LiTan2025} M.-S. Li and Y.-X. Tan, Bargmann invariants for quantum imaginarity, Phys. Rev. A \textbf{111}, 022409 (2025). 

\bibitem{Li2026} M.-S. Li, R. Wagner, and L. Zhang, Multistate imaginarity and coherence in qubit systems, Phys. Rev. A \textbf{113}, 012428 (2026). 

\bibitem{WuPRL2021} K.-D. Wu, T. V. Kondra, S. Rana, C. M. Scandolo, G.-Y. Xiang, C.-F. Li, G.-C. Guo, and A. Streltsov, Operational resource theory of imaginarity, Phys. Rev. Lett. \textbf{126}, 090401 (2021). 

\bibitem{Wu2024} K.-D. Wu, T. V. Kondra, C. M. Scandolo, S. Rana, G.-Y. Xiang, C.-F. Li, G.-C. Guo, and A. Streltsov, Resource theory of imaginarity in distributed scenarios, Commun. Phys. \textbf{7}, 171 (2024). 

\bibitem{Xue2021} S. Xue, J. Guo, P. Li, M. Ye, and Y. Li, Quantification of resource theory of imaginarity, Quantum Inf. Process. \textbf{20}, 383 (2021). 

\bibitem{Xu2024} J. Xu, Quantifying the imaginarity of quantum states via Tsallis relative entropy, Phys. Lett. A \textbf{528}, 130024 (2024). 

\bibitem{DuBai2025} S. Du and Z. Bai, Quantifying imaginarity in terms of pure-state imaginarity, Phys. Rev. A \textbf{111}, 022405 (2025). 

\bibitem{Guan2026} N.-N. Guan, M.-L. Guo, X.-N. Zhu, Z.-X. Jin, and Y.-H. Zhou, A convex-roof affinity-square measure of quantum imaginarity, Adv. Quantum Technol. \textbf{9}, e70513 (2026). 

\bibitem{Kondra2023} T. V. Kondra, C. Datta, and A. Streltsov, Real quantum operations and state transformations, New J. Phys. \textbf{25}, 093043 (2023). 

\bibitem{Guo2025} M.-L. Guo, B. Li, and S.-M. Fei, Geometric-like imaginarity: Quantification and state conversion, Sci. China Phys. Mech. Astron. \textbf{68}, 220311 (2025). 

\bibitem{Petz1996} D. Petz, Monotone metrics on matrix spaces, Linear Algebra Appl. \textbf{244}, 81 (1996). 

\bibitem{Braunstein1994} S. L. Braunstein and C. M. Caves, Statistical distance and the geometry of quantum states, Phys. Rev. Lett. \textbf{72}, 3439 (1994). 

\bibitem{Lu2026} X. Lu, Z. Guo, H. Cao, and Y. Fan, Quantum-Fisher-information-based imaginarity quantifiers under covariant real operations, Phys. Rev. A (accepted 13 November 2026). 

\bibitem{Hayashi2018Comparison} A. Hayashi, T. Hashimoto, and M. Horibe, Quantum-state comparison and discrimination, Phys. Rev. A \textbf{97}, 052323 (2018). 

\bibitem{Gour2020} G. Gour and M. Tomamichel, Optimal extensions of resource measures and their applications, Phys. Rev. A \textbf{102}, 062401 (2020). 

\bibitem{Vidal2000} G. Vidal, Entanglement monotones, J. Mod. Opt. \textbf{47}, 355 (2000). 

\bibitem{Baumgratz2014} T. Baumgratz, M. Cramer, and M. B. Plenio, Quantifying coherence, Phys. Rev. Lett. \textbf{113}, 140401 (2014). 

\bibitem{Streltsov2017} A. Streltsov, G. Adesso, and M. B. Plenio, Colloquium: Quantum coherence as a resource, Rev. Mod. Phys. \textbf{89}, 041003 (2017). 

\bibitem{VidalTarrach1999} G. Vidal and R. Tarrach, Robustness of entanglement, Phys. Rev. A \textbf{59}, 141 (1999). 

\bibitem{Steiner2003} M. Steiner, Generalized robustness of entanglement, Phys. Rev. A \textbf{67}, 054305 (2003). 

\bibitem{Napoli2016} C. Napoli, T. R. Bromley, M. Cianciaruso, M. Piani, N. Johnston, and G. Adesso, Robustness of coherence: An operational and observable measure of quantum coherence, Phys. Rev. Lett. \textbf{116}, 150502 (2016). 

\bibitem{Regula2018} B. Regula, Convex geometry of quantum resource quantification, J. Phys. A: Math. Theor. \textbf{51}, 045303 (2018). 

\bibitem{TakagiRegula2019} R. Takagi and B. Regula, General resource theories in quantum mechanics and beyond: Operational characterization via discrimination tasks, Phys. Rev. X \textbf{9}, 031053 (2019). 

\bibitem{Helstrom1976} C. W. Helstrom, \textit{Quantum Detection and Estimation Theory} (Academic Press, New York, 1976).

\bibitem{XieEberly2021} S. Xie and J. H. Eberly, Triangle measure of tripartite entanglement, Phys. Rev. Lett. \textbf{127}, 040403 (2021). 

\bibitem{Ge2023} X. Ge, L. Liu, and S. Cheng, Tripartite entanglement measure under local operations and classical communication, Phys. Rev. A \textbf{107}, 032405 (2023). 

\bibitem{Boyd2004} S. Boyd and L. Vandenberghe, \textit{Convex Optimization} (Cambridge University Press, Cambridge, 2004). 

\bibitem{Magni2026} B. Magni, M. Heinrich, L. Leone, and X. Turkeshi, Anticoncentration and state design of doped real Clifford circuits and tensor networks, Phys. Rev. A \textbf{113}, 062446 (2026). 

\bibitem{Temme2010} K. Temme, M. J. Kastoryano, M. B. Ruskai, M. M. Wolf, and F. Verstraete, The $\chi^2$-divergence and mixing times of quantum Markov processes, J. Math. Phys. \textbf{51}, 122201 (2010). 

\bibitem{Huebner1992} M. H{\"u}bner, Explicit computation of the Bures distance for density matrices, Phys. Lett. A \textbf{163}, 239 (1992). 

\bibitem{Dittmann1999} J. Dittmann, Explicit formulae for the Bures metric, J. Phys. A: Math. Gen. \textbf{32}, 2663 (1999). 

\bibitem{Uhlmann1976} A. Uhlmann, The ``transition probability'' in the state space of a $*$-algebra, Rep. Math. Phys. \textbf{9}, 273 (1976). 

\bibitem{Liu2026} J. Liu, Y. Sun, J. Xu, and M.-J. Zhao, Imaginarity measures induced by real part states and the complementarity relations, J. Phys. A: Math. Theor. \textbf{59}, 035302 (2026). 

\bibitem{HornJohnson2013} R. A. Horn and C. R. Johnson, \textit{Matrix Analysis}, 2nd ed. (Cambridge University Press, Cambridge, 2013). 

\bibitem{Holevo1982} A. S. Holevo, \textit{Probabilistic and Statistical Aspects of Quantum Theory} (North-Holland, Amsterdam, 1982).

\bibitem{Paris2009} M. G. A. Paris, Quantum estimation for quantum technology, Int. J. Quantum Inf. \textbf{7}, 125 (2009). 

\bibitem{LiuQFI2020} J. Liu, H. Yuan, X.-M. Lu, and X. Wang, Quantum Fisher information matrix and multiparameter estimation, J. Phys. A: Math. Theor. \textbf{53}, 023001 (2020). 

\bibitem{Zhu2021} H. Zhu, Hiding and masking quantum information in complex and real quantum mechanics, Phys. Rev. Research \textbf{3}, 033176 (2021). 

\bibitem{ZhangBroadcast2024} Z. Zhang, N. Li, and S. Luo, Broadcasting of imaginarity, Phys. Rev. A \textbf{110}, 052439 (2024). 

\bibitem{Sun2025} Y. Sun, R. Ren, Y. Wang, and Y. Li, Analysis of the relationship between imaginarity and entanglement, Phys. Rev. A \textbf{111}, 032425 (2025). 

\bibitem{Biswas2026} I. Biswas, S. Bera, U. Sen, I. Chattopadhyay, and D. Sarkar, Increasing entanglement concentration and the role of measurement imaginarity, Phys. Rev. A \textbf{113}, 062439 (2026). 

\bibitem{ChenLei2025} X. Chen and Q. Lei, Imaginarity of quantum channels: Refinement and alternative, Phys. Lett. A \textbf{530}, 130129 (2025). 

\bibitem{Rockafellar1970} R. T. Rockafellar, \textit{Convex Analysis} (Princeton University Press, Princeton, NJ, 1970).
\end{thebibliography}
\end{document}